\documentclass{aa}

\usepackage{graphicx}
\usepackage{xcolor}
\usepackage{subfig}
\usepackage{txfonts}
\usepackage{booktabs}
\usepackage[
    colorlinks=true,    
    linkcolor=blue,     
    urlcolor=cyan,      
    citecolor=blue,      
    bookmarks=true,     
    pdftitle={CtaAgnVar: a Pipeline for Variability Studies of Blazars with the Cherenkov Telescope Array Observatory},  
    pdfauthor={Guillaume GROLLERON}
]{hyperref}
\begin{document}

   \title{\textsc{CtaAgnVar}: a Pipeline for Variability Studies with the Cherenkov Telescope Array Observatory}


   \author{G.~Grolleron\inst{\ref{lapp}}
   \and
   J.-P.~Lenain\inst{\ref{lpnhe}}
   \and
   J.~Biteau\inst{\ref{ijclab},~\ref{iuf}}
   \and
   C.~Boisson\inst{\ref{luth}}
   \and
   M.~Cerruti\inst{\ref{apc}}
}
   \institute{
    Univ. Savoie Mont Blanc, CNRS, Laboratoire d'Annecy de Physique des Particules - IN2P3, 76000 Annecy, France\label{lapp}
    \and
    Sorbonne Universit\'e, CNRS/IN2P3, Laboratoire de Physique Nucl\'eaire et de Hautes Energies, LPNHE, 4 place Jussieu, 75005 Paris, France \label{lpnhe}
    \and
    Université Paris-Saclay, CNRS/IN2P3, IJCLab, Orsay, France \label{ijclab}
    \and 
    Institut universitaire de France (IUF), France \label{iuf}
    \and
    LUX,  Observatoire de Paris, Université PSL, CNRS, Sorbonne Université, 5 Pl. Jules Janssen, 92190 Meudon, France \label{luth}
    \and
    APC, Université Paris Cité, Paris, France \label{apc}
   }

   \date{}

 
  \abstract{Blazars are active galactic nuclei (AGN) with a relativistic jet oriented toward the observer. This jet accelerates particles to relativistic energies, which subsequently radiate over the entire electromagnetic spectrum. Flux and spectral variability have been observed on short- and long-term time scales in blazars. Its characterization allows us to understand the underlying physical processes and then to constrain the modeling of the central black hole and its jets. The study of blazars is therefore one of the key science cases of the Cherenkov Telescope Array Observatory (CTAO). The CTAO constitutes the next generation of Imaging Atmospheric Cherenkov Telescope (IACT). It will be composed of two sites, one per hemisphere. The CTAO will be more sensitive by a factor of five to ten depending on the energy than current IACTs. Therefore, we expect that the reconstruction of AGN variability at very high gamma-ray energies will achieve a level of precision never reached before. In the context of developing the CTAO science cases, simulations are performed to forecast the capabilities of the CTAO. Concerning blazars and more generally transient phenomena, to quantify the CTAO capabilities, we developed \textsc{CtaAgnVar}, a tool based on \textsc{Gammapy}, the open-source Python package for gamma-ray analysis selected as core library for the Science Analysis Tool of the CTAO, to perform simulations and analyses of blazar variability. This pipeline covers the full chain from a time-dependent spectral energy distribution to the simulation of a realistic observation sequence, accounting for the visibility of the source, and to the reconstruction of spectra and light-curves. It provides a set of variability estimators, among which a statistical estimator, introduced in this work, that quantifies the detection of hysteresis patterns in hardness ratio diagrams. \textsc{CtaAgnVar} is released as open-source software and the instrument enters the pipeline only through the instrument response functions and the location of the array, so that its use is not restricted to the CTAO. Its functionalities, analysis modes and configurable options are summarized in a dedicated table. To demonstrate the scientific merit of the pipeline, we simulated over 15 years of CTAO observations of PG 1553$+$113 with the CTAO to forecast the perspectives on the periodicity detection of this source at very high energies. In \textit{Fermi}-LAT data, this source presents a periodicity in high-energy gamma rays and we studied to what extent the CTAO could reconstruct such a periodicity in very high-energy gamma rays. We show that, under the assumption that the very high-energy emission follows the high-energy one, the CTAO would detect the same gamma-ray emission periodicity as \textit{Fermi}-LAT, with a significance level above $5~\sigma$, with 15 years of 30 min observations taken on a weekly cadence, the determination of the period being then limited by the sampling of the light-curve rather than by the photon statistics.}

   \keywords{Active galactic nuclei, High energy astrophysics, Gamma rays
}

\maketitle
\nolinenumbers
%

\section{Introduction}
\label{section 1}
Relativistic jets launched by supermassive black holes are among the most extreme particle accelerators in the Universe. When the jet of an active galactic nucleus (AGN) is closely aligned with our line of sight, the object is classified as a blazar. These sources emit radiation across the entire electromagnetic spectrum, from radio waves to very high-energy (VHE, $E>100~\mathrm{GeV}$) gamma rays~\citep{2019NewAR..8701541H}. They are the most numerous type of source in the extragalactic high-energy (HE, $E>100~\mathrm{MeV}$) gamma-ray sky~\citep{2022ApJS..260...53A,2020ApJ...892..105A}. This broadband emission is produced by particles accelerated within the relativistic jets~\citep{1979ApJ...232...34B}. In leptonic models, charged particles radiate via synchrotron emission, which can then be up-scattered through the inverse Compton process on the synchrotron photons themselves or on external photon fields from the accretion disk, the broad-line region, or the dusty torus. If the accelerated particles are hadrons, their interaction processes can also produce VHE gamma rays. The complexity of blazar emission thus provides a probe of both the central engine properties, such as the black hole mass and accretion rate~\citep{2002MNRAS.332..231U,2006Natur.444..730M}, and the physical conditions within the jet, such as the magnetic field and the bulk Lorentz factor~\citep{2001A&A...367..809K,2008A&A...478..111L}.\\

Blazars exhibit temporal variability in both their flux and spectral properties on a wide range of timescales, from minutes-long flares to year-long variations~\citep{2010A&A...520A..83H,2016ApJ...824L..20A,2019NewAR..8701541H}. The characterization of this variability provides a unique window into the physics of blazars, allowing us to study the properties of the jet and the central black hole, and to discriminate between different emission models. 
Observations of blazars will be a cornerstone of the key science programs for the upcoming Cherenkov Telescope Array Observatory~\citep[CTAO,][]{2013APh....43..215S,SciencewithCTA}. In this context, we aim to quantify the future capabilities of the CTAO to provide new insights into blazars, particularly concerning time-dependent phenomena. With current Imaging Atmospheric Cherenkov Telescopes (IACTs), reconstructing light-curves with a time resolution better than a day is extremely challenging and has been achieved for only a few bright sources at VHE, such as Mrk 421~\citep{2020ApJS..248...29A}, Mrk 501~\citep{2007ApJ...669..862A}, and PKS~2155$-$304~\citep{2007ApJ...664L..71A}. This short list illustrates where the current generation stands: sub-hour variability is within reach only during exceptional flares of the brightest sources, and the associated spectra can rarely be resolved in time. From a phenomenological standpoint, exploring fast variability is crucial for understanding the emission mechanisms. The rise and fall times of flares are directly related to the acceleration and cooling timescales of the underlying population of radiating particles, as well as on the size of the emitting region and the physical properties of the medium~\citep{2008ApJ...686..181F}. Furthermore, observing whether flares are achromatic (affecting all energies equally) or chromatic (affecting different energy bands differently) can reveal the balance between the processes that dominate acceleration and cooling timescales~\citep{2020ApJS..248...29A}. In multi-wavelength observations, the sub-night time resolution achievable in the optical, UV, or X-ray bands is much finer than what is feasible due to sensitivity limits of the current generation of IACTs, limiting our ability to constrain multi-wavelength models. Finally, characterizing the long-term behavior of AGN and their duty cycles (the ratio of active to quiescent time) remains challenging but is critical for understanding their emission mechanisms~\citep{2006Natur.444..730M}.

The future CTAO, which will be five to ten times more sensitive than current-generation IACTs depending on the energy, will significantly improve the temporal resolution of AGN variability studies, helping to overcome the aforementioned limitations. The CTAO will consist of two sites, one in each hemisphere. The southern site will be located in Chile, and the northern site at La Palma, Spain. This configuration maximizes the observable sky. It is important to remember that to observe AGN with IACTs a relatively low energy threshold is needed (except for bright or nearby sources). Because of the moonlight contamination at low energy, such a low threshold can only be obtained with moonless observations, inducing a reduction of the available periods where AGN observations can be scheduled. The CTAO will feature three main types of telescopes: Large-Sized Telescopes (LSTs), Medium-Sized Telescopes (MSTs), and Small-Sized Telescopes (SSTs). LSTs are most sensitive to the lower end of the VHE range (tens of GeV), while SSTs are optimized for the highest energies (tens of TeV). MSTs cover the intermediate energy range. The northern site array, focused on extragalactic studies, will be composed of LSTs and MSTs, whereas the southern one will be additionally composed of SSTs to cover the Galactic science\footnote{4 LSTs and 9 MSTs are expected in the northern site. While the southern site will be composed of 2 LSTs, 14 MSTs and 37 SSTs.}.

\textsc{Gammapy}~\citep{gammapy:2023} is the official software for the CTAO science tools. This package allows for the analysis of data from existing IACTs like H.E.S.S., MAGIC, and VERITAS, and also provides support for \textit{Fermi}-LAT and HAWC data. With \textsc{Gammapy}, it is also possible to simulate CTAO observations based on its expected instrument response functions ~\citep[IRFs]{cherenkov_telescope_array_observatory_2021_5499840}, which are derived from Monte Carlo (MC) simulations of the full detection chain: the development of air showers in the atmosphere and the associated Cherenkov light emission, simulated with \textsc{CORSIKA}~\citep{1998cmcc.book.....H}, followed by the reflection of the Cherenkov photons on the telescope mirrors and the response of the photosensors and of the readout electronics, simulated with \textsc{sim\_telarray}~\citep{2008APh....30..149B}. However, while efforts have been made in this direction, the library was not initially designed to fully support time-dependent simulations. We have therefore developed a pipeline, \textsc{CtaAgnVar}, as a \textsc{Gammapy} overlay to perform simulations and reconstructions of time-dependent phenomena. The package was initially created to study AGN variability, but its use can be extended to any type of time-dependent astrophysical source emitting in the VHE range. The \textsc{CtaAgnVar} pipeline is now being extensively used for establishing the CTAO AGN key science project~\citep{2023arXiv230909615C,2023arXiv230912157G}. It allows for realistic simulations of blazar flares and long-term behavior with the CTAO, using any kind of blazar emission model. Moreover, \textsc{CtaAgnVar} is not tied to a specific IRF and can be used with any CTAO IRFs (e.g., for sub-arrays simulations). The pipeline is released as open-source software under a BSD 3-clause license\footnote{\url{https://gitlab.cta-observatory.org/guillaume.grolleron/ctaagnvar}}, so that it can also be used by the community beyond the CTAO Consortium. In this article, we explain the workflow of \textsc{CtaAgnVar}, from the input model to the simulation of gamma-like events and the reconstruction of spectra and light-curves. The pipeline itself is the main result presented here: its functionalities, analysis modes and configurable options are summarized in Table~\ref{tab:functionalities}. The second, equally important part of this work is scientific: the assessment presented in Sec.~\ref{section 5} of the CTAO capability to detect at VHE the periodicity of PG~1553$+$113, which provides a direct input for the definition of AGN monitoring strategy of the CTAO to search for periodicity at VHE.

The search for periodic emission is one of the key time-dependent phenomena investigated in blazar observations. Such periodicity could be linked to the orbital motion in a binary supermassive black hole system, jet precession, or other rotational effects in the central engine~\citep{2004ApJ...615L...5R}. Periodicity has been detected in \textit{Fermi}-LAT data for five AGN with a significance above 5 sigma~\citep{2022arXiv221101894P}. PG~1553$+$113 is the blazar with the most significant detection. Its emission periodicity was already claimed in~\cite{2015ApJ...813L..41A} but with a lower significance. At VHE, no periodicity has been detected so far with a decade of MAGIC observations~\citep{2024MNRAS.529.3894M}. From a more general point of view, as discussed in~\cite{2023A&A...672A..86R}, the search for periodicity in blazar emission is a challenging topic mostly because of two points. First, an unbiased time series is needed, this can be easily achieved with \textit{Fermi}-LAT but it is strongly more difficult with IACTs because observations are performed more frequently during high states, rather than in quiescent states. Secondly, because of the large number of scales probed, it is crucial to apply trial corrections to avoid any false positive.

Building on the capabilities of \textsc{CtaAgnVar}, we have simulated observations of PG~1553$+$113 with the CTAO to investigate to what extent the CTAO could detect this periodicity at VHE, and which observation strategy would be required to do so.

The input required by \textsc{CtaAgnVar} and the simulation setup are explained in Sec.~\ref{section 2}. In Sec.~\ref{section 3}, we detail the simulation of gamma-like events and the computation of the source visibility. Then, the high-level analysis, namely the reconstruction of the spectrum and light-curve, is described in Sec.~\ref{section 4}. An application to a blazar emission model, based on \textit{Fermi}-LAT data, to study the periodicity of PG~1553$+$113 with the CTAO is presented in Sec.~\ref{section 5}, before concluding in Sec.~\ref{conclusion}.

\section{Injected models and simulation set-up}
\label{section 2}

\subsection{Overview of the pipeline}
\label{subsec:overview}
As mentioned, the goal of \textsc{CtaAgnVar} pipeline is to provide realistic simulations of CTAO observations. Those simulations are then provided to the CTAO extragalactic working group to assess to what extent the CTAO would be able to reconstruct the injected model, and to discriminate between some emission models (e.g. hadronic or leptonic).

The pipeline is organized as a linear workflow, each step of which is configurable. A time-dependent spectral energy distribution (SED) is provided as input, either as an analytic formula or as a time series of SED snapshots (Sec.~\ref{subsec:injected}). An observation sequence is then built from the visibility of the source, the observation duration and the cadence requested by the user, and one set of IRFs is associated to each observation according to the zenith angle and to the night sky background level (Sec.~\ref{subsec:setup}). Gamma-like events are subsequently simulated in each observation as Poisson realizations of the expected counts (Sec.~\ref{section 3}). The resulting datasets are analyzed to reconstruct spectra and light-curves, with an automatic selection of the spectral hypothesis and, if requested, an adaptive time binning. Finally, a set of estimators is applied to the reconstructed light-curves to characterize the variability, from the fractional variance to the search for spectral hysteresis and for periodicity (Sec.~\ref{section 4}).

This modularity translates into a variety of use cases. Short observations with a fine time sampling are used to study the fast variability and the spectral evolution during bright flares, whereas long exposures with a low cadence are suited to long-term monitoring and to the search for periodicity, which is the subject of the study presented in Sec.~\ref{section 5}. Simulating competing emission models with the same observation sequence quantifies the ability of the CTAO to discriminate between them, for instance between leptonic and hadronic scenarios. Since the pipeline is not tied to blazars, it is also applicable to other time-dependent phenomena, such as gamma-ray bursts, and to studies of the violation of Lorentz invariance. It can finally be used to analyze real data, and not only simulated ones. The functionalities, the analysis modes and the main configurable options are summarized in Table~\ref{tab:functionalities}, with a pointer to the section where each of them is described.

\begin{table*}[t]
\caption{Main functionalities, analysis modes and configurable options of \textsc{CtaAgnVar}.}
\label{tab:functionalities}
\centering
\small
\begin{tabular}{p{0.16\textwidth}p{0.33\textwidth}p{0.37\textwidth}c}
\toprule
\textbf{Stage} & \textbf{Functionality} & \textbf{Main configurable options} & \textbf{Sect.} \\
\midrule
Input model & Time-dependent SED, with arbitrary and possibly non-uniform time binning & Analytic formula, time series of SED snapshots, stationary source, or model built from the \textit{Fermi}-LAT LCR; EBL absorption included in the input model or applied by the pipeline; redshift & \ref{subsec:injected} \\
\addlinespace
Observation sequence & Computation of the darkness and visibility windows, scheduling of the observations & Site (North or South); observation live time; cadence; starting time and total duration; allowed zenith-angle range; moon-dependent sky quality; alternatively, a user-provided list of time bins bypassing the visibility computation & \ref{subsec:setup} \\
\addlinespace
Instrument response & Selection of one set of IRFs per observation & Any CTAO IRF set, including sub-arrays (LSTs, MSTs, SSTs and their combinations); IRFs at 20, 40 and 60 degrees of zenith distance; standard or increased night sky background; fixed IRF mode without source tracking & \ref{subsec:setup} \\
\addlinespace
Event simulation & Poisson realizations of the expected counts in the ON and OFF regions & Number of realizations; energy range and number of energy bins; ON region radius; offset & \ref{sec:simu} \\
\addlinespace
Spectral reconstruction & Aperture photometry likelihood fit (WStat) maximized with \textsc{Minuit} & Any spectral model implemented in \textsc{Gammapy}, in particular Eqs.~\ref{eq:pl} to~\ref{eq:eclp}; EBL-absorbed models; each parameter free or fixed, including the redshift; goodness-of-fit $p$-value; safe energy range from a minimum number of counts per energy bin & \ref{subsec:spectrum} \\
\addlinespace
Light-curve reconstruction & Spectral fit in each time bin with an automatic selection of the spectral hypothesis & Significance threshold of the likelihood ratio test used for the model selection; constant or adaptive time binning driven by a target significance per bin; light-curves in energy sub-bands; masking of the bins below a given significance & \ref{Procedure} \\
\addlinespace
Variability estimators & Fractional variability of the flux and of the photon index, point-to-point fractional variability & Choice of the reconstructed quantity; light-curve used as input & \ref{subsec:fvar} \\
\addlinespace
Spectral variability & Hardness ratio diagrams and detection of hysteresis patterns with the estimator $T$ & Energy boundaries defining the two bands; number of toy MC realizations used for the $H_0$ and $H_1$ hypotheses & \ref{subsec:HR} \\
\addlinespace
Periodicity search & Lomb-Scargle periodogram with a bootstrap false alarm probability accounting for trial factors & Range of periods probed; number of bootstrap realizations & \ref{subsec:periodic_search} \\
\bottomrule
\end{tabular}
\end{table*}

\subsection{Injected models}
\label{subsec:injected}
To perform simulations of blazar observations (or other kind of sources) with the CTAO, a spectral model of the source emission is needed. Essentially, it consists in a time-dependent SED with a given time span and time resolution. The time bins do not need to be identical. Hence, \textsc{CtaAgnVar} can use as input for the source modelling both a basic analytic formula describing the time-dependent SED evolution, or the SED time series resulting from a time-dependent modeling framework, for instance the numerical solution of a Fokker-Planck equation describing the evolution of the radiating particle distribution~\citep[e.g.,][]{2022A&A...658A.173T}.

When a VHE gamma-ray travels in the Universe, it can be absorbed by the extragalactic background light (EBL) which is the starlight emission and its re-processing by dust in galaxies in the whole Universe. It induces an absorption in the VHE spectrum due to $\gamma$-$\gamma$ interactions. This effect can be included in the injected model, or it can alternatively be added directly within \textsc{CtaAgnVar}. In such a case, the EBL model used is the one from~\citet{2011MNRAS.410.2556D}.

\subsection{Observation set-up and source visibility}
\label{subsec:setup}
Several parameters can be configured to tune the simulations. The main ones specify the source that is observed or the duration of observations, as well as the cadence.
An observation is defined as a continuous data acquisition. Usually, the duration, or live time, of an observation is between 20 and 30 min with IACTs. During a night, \textsc{CtaAgnVar} simulates several observations as long as the observed source is visible under good conditions. This can be defined regarding the required darkness level or the maximum source zenith angle allowed\footnote{Observations at large zenith angles lead to a higher energy threshold, which is limiting for extragalactic studies such as AGN observations.}.

Cadence is the frequency at which a source is observed. For instance, capturing the light curve of a bright, flaring AGN requires the highest possible time resolution, and thus a high observation cadence. On the other hand, long-term monitoring often observes a source in relatively low states; therefore, to optimize a given time budget, it may be better to use longer exposures with a lower (e.g., weekly or monthly) cadence. While an upcoming CTAO Consortium publication will address these observation strategies in detail, we note here that the \textsc{CtaAgnVar} pipeline is designed to handle various combinations of cadence and exposure. 

The input model needs to start at a given time reference, which can be specified by the user. This moment will be used thereafter to compute the visibility of the source. The CTAO observing site can be selected and the telescope configurations as well, hence it is possible to select LSTs only or MSTs only for example. The spectral model used to fit the data and reconstruct a spectrum or a light-curve can then be tuned. It can be selected among analytic models (power-law, with or without an energy cut-off, log-parabola, etc) or any model implemented within \textsc{Gammapy}. 

To simulate realistic observations, the tracking of the source is accounted for, thus its visibility along the night is computed. We use public IRFs of the CTAO~\citep{cherenkov_telescope_array_observatory_2021_5499840}. They are produced with MC simulations performed with pointing with a zenith distance of 20, 40 and 60 degrees. In \textsc{CtaAgnVar}, IRFs are thus selected according to the zenith angle evolution along the night. The 20° IRF is used when the source is below 25°, the 40° IRF is used when the source is between 25° and 45°, and the 60° IRF is used when the source is between 45 and 65°. Observations with a zenith distance larger than 65° are excluded. The computation of the source position is performed with the python package \textsc{PyEphem}~\citep{2011ascl.soft12014R}. \\

First, from the starting time specified in the setup, the next darkness period is found, then the visibility period is computed for the night, with a criterion on the allowed range of zenith angles that can be specified by the user. Time bins for the observations are then created following the cadence and the live time of each observation specified by the user.

Next, knowing that the injected model does not have the same time binning as the simulated one, it is interpolated in time and the interpolated SED is selected for each observation time bins within the night. The same procedure is done to the following night until the specified time duration is reached. In parallel, IRFs are selected following the source position.\\
This computation of the visibility period with the source tracking can be bypassed. Users can provide directly a list of times bins when they assume the source to be visible, if so, the observations sequence will be created following these bins and the source zenith angle will be determined to properly select IRFs. Furthermore, a sky quality parameter may be provided associated to time bins to make possible the use of both CTAO IRFs, with a standard night sky background (NSB) and an increased one, which corresponds to moonlight period with moderate illumination. This sky quality takes into account the angular distance between the observed source and the moon and its luminosity according to the moon's cycle.

At this step, the observation sequence has been set up with one injected model and one IRF per time bin. \textsc{Gammapy} datasets are then created, these are data structures associated to observations. One dataset is linked to one observation as described above, and within datasets, photon counts are simulated.

\section{Simulations and Analysis}
\label{section 3}

\subsection{Simulations of gamma-like events}
\label{sec:simu}

Cherenkov astronomy relies on the reconstruction of the energy, direction, and nature of the primary particles from the camera images of the air showers that they induce in the atmosphere. Cosmic rays are far more numerous than gamma rays and produce air showers whose images partly resemble the gamma-ray ones, so that the two populations must be discriminated at the reconstruction level. Such air shower simulations are not performed here: they are carried out at a lower stage in the data flow and are used to produce the IRFs. In \textsc{Gammapy}, only gamma-like events detected by the instrument are simulated from those IRFs. Essentially the number of expected counts due to the source can be known from the SED and the convolution with the IRFs. The background counts can be known from MC simulations of air showers induced by cosmic rays, depending on the NSB level, which are encoded within the IRFs. The predicted total count follows a Poisson law, therefore the simulated counts are simply obtained with a random draw following this law. The process is performed for each time step. Thus, at this stage the gamma-like events have been simulated for the list of datasets. The data can then be analyzed to produce high level data products, such as spectra and light-curves. 

Although studies of sources with complex fields, such as Galactic ones, require 3D analysis based on likelihood computation with proper background and source modeling~\citep{1979ApJ...228..939C}, extragalactic sources, expected to be point-like, can simply be analyzed using aperture photometry (or 1D) analysis in gamma-ray astronomy. The procedure is described in \citet{1983ApJ...272..317L,2007A&A...466.1219B}. First, a region around the expected position of the source is defined, this region is usually called ON region. OFF regions are defined where the background will be evaluated. Therefore, the expected gamma-like events produced by the source, $N_{\mathrm{excess}}$, is defined as follows :
\begin{equation}
    N_{\mathrm{excess}} = N_{\mathrm{ON}} - \alpha N_{\mathrm{OFF}}
\end{equation}
with $\alpha$ the ratio between the ON region area and the OFF region one. 
This analysis is standard and it has been proven that for point-source observations, which is the case for blazars at VHE, sophisticated 3D analyses give comparable detection significance~\citep{2007A&A...466.1219B,grolleron:tel-04843969}.

\subsection{Reconstruction of spectral properties}
\label{subsec:spectrum}

To reconstruct spectral properties, a model is fitted to the data. Any of the spectral models implemented in \textsc{Gammapy} can be used for this purpose. In practice, the VHE spectra of blazars are described by a small set of physically motivated parameterizations, defined in Eqs.~\ref{eq:pl} to~\ref{eq:eclp}: a power-law (Eq.~\ref{eq:pl}), to which a curvature (Eq.~\ref{eq:lp}) or an exponential cut-off (Eq.~\ref{eq:ecpl}) can be added. The curvature reflects the shape of the underlying particle distribution, while the cut-off is related to the maximum energy reached by the radiating particles. In \textsc{CtaAgnVar}, the EBL can be automatically included in the selected spectral model. The EBL absorption depends on the distance traveled by gamma rays from the source to the observer, directly related to the redshift, and the energy of the gamma rays. The redshift can in principle be left free in the fitting process but for most of the cases it is fixed to the best known value corresponding to the source. All the spectral parameters can be fixed or left free.
The likelihood of the number of excess events predicted by the fitted spectral model is then evaluated on the simulated data $N_\mathrm{excess}$. This is a binned likelihood in energy, defined by the WStat formula, implemented in \textsc{Gammapy}, which comes from \textsc{XSPEC}~\citep{1996ASPC..101...17A}. In this formula, no further assumption is needed for the background. 
\textsc{Minuit}~\citep{James:1975dr,dembinski_2024_10638795} is used to maximize the log-likelihood. 

In order to assess the validity of the reconstructed spectrum, a likelihood ratio test between the data and the model is used. It has been verified that this ratio asymptotically follows a chi-square with $n-m$ degrees of freedom, with $n$ the number of energy bins and $m$ the number of free parameters in the model. Then it is possible to define a $p$-value that is used to reject a hypothesis on the model at some specific confidence level. In \textsc{CtaAgnVar}, a limitation on the minimum number of counts per energy bin has been implemented to set a maximum energy which is provided by the code.

\subsection{Light-curve reconstruction}
\label{Procedure}

To reconstruct a light-curve, a spectral model is fitted to the simulated data for each time bin, from the simplest spectral hypothesis to the most complex one. It is essentially the same as the process described in the previous paragraph, but here not all the \textsc{gammapy} implemented models can be used. 
First a basic power-law (see Eq. \ref{eq:pl}) and a more complex model are fitted for a given time bin. If the likelihood-ratio test rejects the simpler model (at a $3~\sigma$ level, which can be tuned), the alternative model is selected and the procedure is iterated until the rejection with the likelihood ratio test is not possible anymore. The alternative models are built by adding to the power-law the physically motivated components discussed above, namely a curvature (see Eq. \ref{eq:lp}) or an exponential cut-off (see Eq. \ref{eq:ecpl}); the combination of both gives the equation \ref{eq:eclp}. This procedure allows us to reconstruct the light-curve precisely and as truly as possible at the same time. The spectral models that are used in the light-curve reconstruction process are defined as follows:
\begin{align}
    \phi_{\rm PL}(E)   &= \phi_0 \left(\frac{E}{E_0}\right)^{-\Gamma} 
                        \label{eq:pl} \\[6pt]
    \phi_{\rm LP}(E)   &= \phi_0 \left(\frac{E}{E_0}\right)^{-\Gamma - \beta\log(E/E_0)} 
                        \label{eq:lp} \\[6pt]
    \phi_{\rm ECPL}(E) &= \phi_0 \left(\frac{E}{E_0}\right)^{-\Gamma} 
                        \exp\left(-\frac{E}{E_{\rm cut}}\right) 
                        \label{eq:ecpl} \\[6pt]
    \phi_{\rm ECLP}(E) &= \phi_0 \left(\frac{E}{E_0}\right)^{-\Gamma - \beta\log(E/E_0)} 
                        \exp\left(-\frac{E}{E_{\rm cut}}\right) 
                        \label{eq:eclp}
\end{align}
where $\phi_0$ is the spectral amplitude at the reference energy $E_0$, 
$\Gamma$ is the photon index, $\beta$ is the spectral curvature of the log-parabola, and $E_{\rm cut}$ is the 
cutoff energy of the exponential cutoff.\\

It is also possible to reconstruct the light-curve in energy sub-bands, which is in particular required for the hardness ratio analysis presented in Sec.~\ref{subsec:HR}.
In order to improve the resolution of the light-curve, it is possible to use a non constant time binning to improve the time resolution. In that case, the detection significance is used to adaptively set the duration of the time bins. Time bins are sliced keeping a significance level per bin defined a priori. This makes it possible to divide a bin into multiple slices and then to play with the balance between detection significance per bins and time resolution of the light-curve.

\section{Variability detection}
\label{section 4}

\subsection{Variability detection with flux and photon index excess variance}
\label{subsec:fvar}
In order to detect time variability, estimators have been implemented in the \textsc{CtaAgnVar} pipeline\footnote{Those estimators are now also available in the latest release of \textsc{Gammapy}}, the first variability estimator is the fractional variability $F_{\mathrm{var}}$~\citep{2003MNRAS.345.1271V} based on excess flux variance $\sigma_{\mathrm{XS}}$~\citep{2002ApJ...568..610E}, defined as follows,
\begin{equation}
    \label{eq:Fvar}
     F_{\mathrm{var}} = \sqrt{\frac{\sigma_{\mathrm{XS}}^2}{<x>^2}}
\end{equation}
where $x$ is the flux. The excess variance, $\sigma_{\mathrm{XS}}^2 =\sigma^2 - <\sigma_{\mathrm{err}}^2>$, is the variance of the flux corrected by the mean of the uncertainties $\sigma_{\mathrm{err}}$.
We can also define the same kind of quantity for the photon index variability detection, simply by replacing the flux $x$ by the photon index in Eq.~\ref{eq:Fvar}, we call the quantity $\Gamma_{\mathrm{var}}$. Both quantities are the simplest way to detect if a variability is detected in flux and in the photon index. 
Moreover, to improve the capability to quantify short-scale variability, the point-to-point fractional variability defined in Eq.~3 from \citet{2002ApJ...568..610E} has been implemented. The definition is as follows, 
\begin{equation}
    F_{\mathrm{pp}} = \frac{1}{<x>} \sqrt{\frac{1}{2(N-1)} \sum_{i=0}^{N-1} (x_{i+1} - x_i)^2 -\sigma_{\mathrm{err}}^2}
\end{equation}
where $N$ is the number of points in the light-curve. 

\subsection{Hardness ratio reconstruction}
\label{subsec:HR}

The hardness ratio (HR), defined as the ratio between flux densities in two distinct energy bands, is a standard tool in X-ray astronomy~\citep[e.g.,][]{1994Sci...264.1313F}. This quantity is used to characterize spectral variability, primarily by exploring the correlation between HR and flux (a relationship referred to as an HR diagram\footnote{Not to be confused with Hertzsprung-Russell diagram}). For blazars, this correlation has been investigated as a fingerprint of the dynamics of the particle acceleration and cooling in the jet. At VHE, a "harder-when-brighter" behavior has been observed with the blazar PKS~2155$-$304~\citep{2007ApJ...664L..71A} and is generally assumed. Conversely, a "softer-when-brighter" behavior has been reported at HE~\citep{2017ApJ...847....7B}. Furthermore, while hysteresis patterns in HR diagrams have been reported in the X-ray domain~\citep{2004A&A...424..841R}—typically attributed to differences between acceleration and cooling timescales~\citep{1996ApJ...470L..89T}—such patterns have not yet been significantly detected at VHE~\citep{2017ApJ...834....2A}. Precise exploration of the VHE flux-HR relationship is thus a key objective for the CTAO to improve our understanding of blazar variability. Consequently, we propose a method to quantify the detection of hysteresis within the HR diagram.

This is a typical example of a measurement that requires the sensitivity of the CTAO. Resolving a hysteresis loop indeed requires several spectrally resolved points within a single flare, that is a time-resolved spectroscopy on timescales much shorter than the flare duration itself. With current IACTs, such a sampling has been reached only during exceptional flares of the few brightest sources mentioned in Sec.~\ref{section 1}, which explains why no significant hysteresis has been reported at VHE so far. The gain in sensitivity of the CTAO, combined with the adaptive time binning described in Sec.~\ref{Procedure}, which balances the significance per bin against the time resolution, and with the reconstruction of light-curves in energy sub-bands, makes this kind of study achievable for a much larger sample of sources and flares.

We define the HR as:
\begin{equation}
    \mathrm{HR} = \frac{\phi(E_2,E_3)}{\phi(E_1,E_2)}
\end{equation}
where $E_1 < E_2 < E_3$ represent three energy boundaries and $\phi(E_a, E_b)$ is the integrated flux between these boundaries.
To quantify the presence of a hysteresis loop, we developed a statistical estimator to assess the HR diagram. While initial attempts used principal component analysis, the methods derived in \citet{grolleron:tel-04843969} proved overly complex. We present here a simpler approach. The goal of the estimator is to robustly reject two distinct hypotheses:
\begin{itemize}
    \item $H_0$: There is no loop in the HR diagram; the points are fully correlated.
    \item $H_1$: A loop exists in the HR diagram, but there is no temporal correlation between points along the pattern (i.e., the sequence is random).
\end{itemize}
Rejecting both $H_0$ and $H_1$ allows us to assert the detection of a temporally coherent hysteresis loop. In the following, the data are represented by two variables, $X$ and $Y$:
\begin{align}
    X &= \{x_j=\ln \phi_j(E_1,E_3),  0 \leq j < N \},\nonumber \\
    Y &= \{y_j=\mathrm{HR}_j, 0 \leq j < N \}
\end{align}
where $N$ is the number of bins in the light-curve.

First, the coordinates $(x_j,y_j)$ are transformed into a polar coordinate system centered on the mean $(\bar{X}, \bar{Y})$, yielding the values $(r_j, \theta_j)$. In this frame, the statistical estimator $T$, directly related to the Shoelace formula~\citep{Braden01091986}, is defined as:
\begin{equation}
    T(X,Y) = \sum_{j=0}^{N-1} \frac{1}{2}|r_j r_{j+1} \sin(\theta_{j+1} - \theta_{j})|
\end{equation}
Each term of this sum is the area of the triangle spanned by the barycenter $(\bar{X}, \bar{Y})$ and two points of the HR diagram that are consecutive in time. The estimator $T$ therefore quantifies the area swept in the HR diagram when following the temporal sequence of the light-curve, without any assumption on the shape of the pattern. Both the extension of the pattern around its barycenter and the temporal ordering of the points contribute to $T$, which is precisely what allows a temporally coherent hysteresis loop to be distinguished from the two hypotheses $H_0$ and $H_1$ defined above. We note that $X$ and $Y$ are not rescaled by their respective standard deviations. The logarithm of the flux and the HR are both dimensionless and span comparable dynamic ranges. More importantly, the value of $T$ is never interpreted on its own: significances are derived by comparing $T$ to the distributions obtained from MC realizations of the same data set, as described below, so that any global rescaling of $X$ and $Y$ leaves the results unchanged.

\begin{figure*}[t]
    \centering
    \subfloat[HR diagram with injected hysteresis]{
        \includegraphics[width=0.32\textwidth]{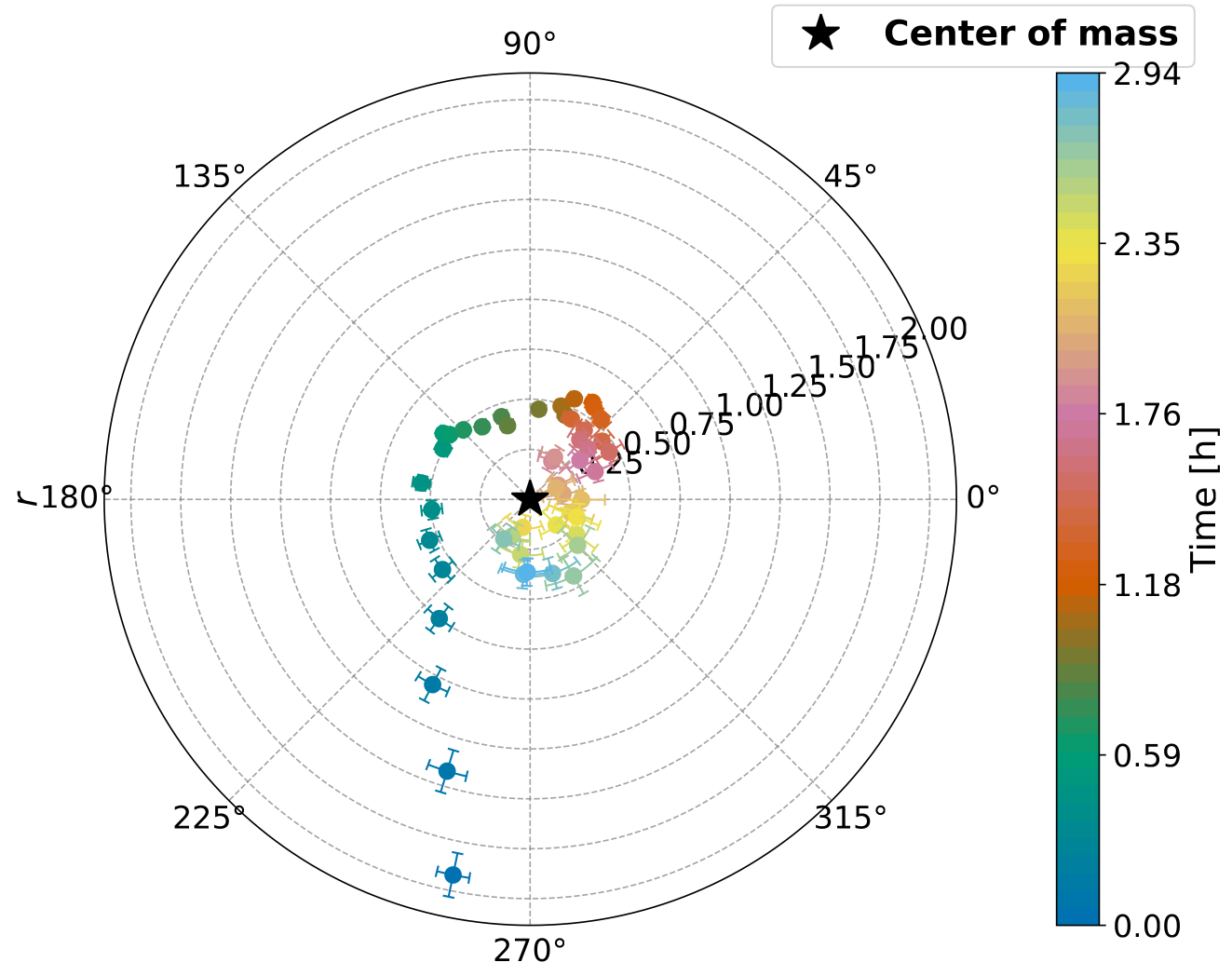}
        \label{fig:hr}
    }
    \subfloat[HR diagram following $H_0$]{
        \includegraphics[width=0.32\textwidth]{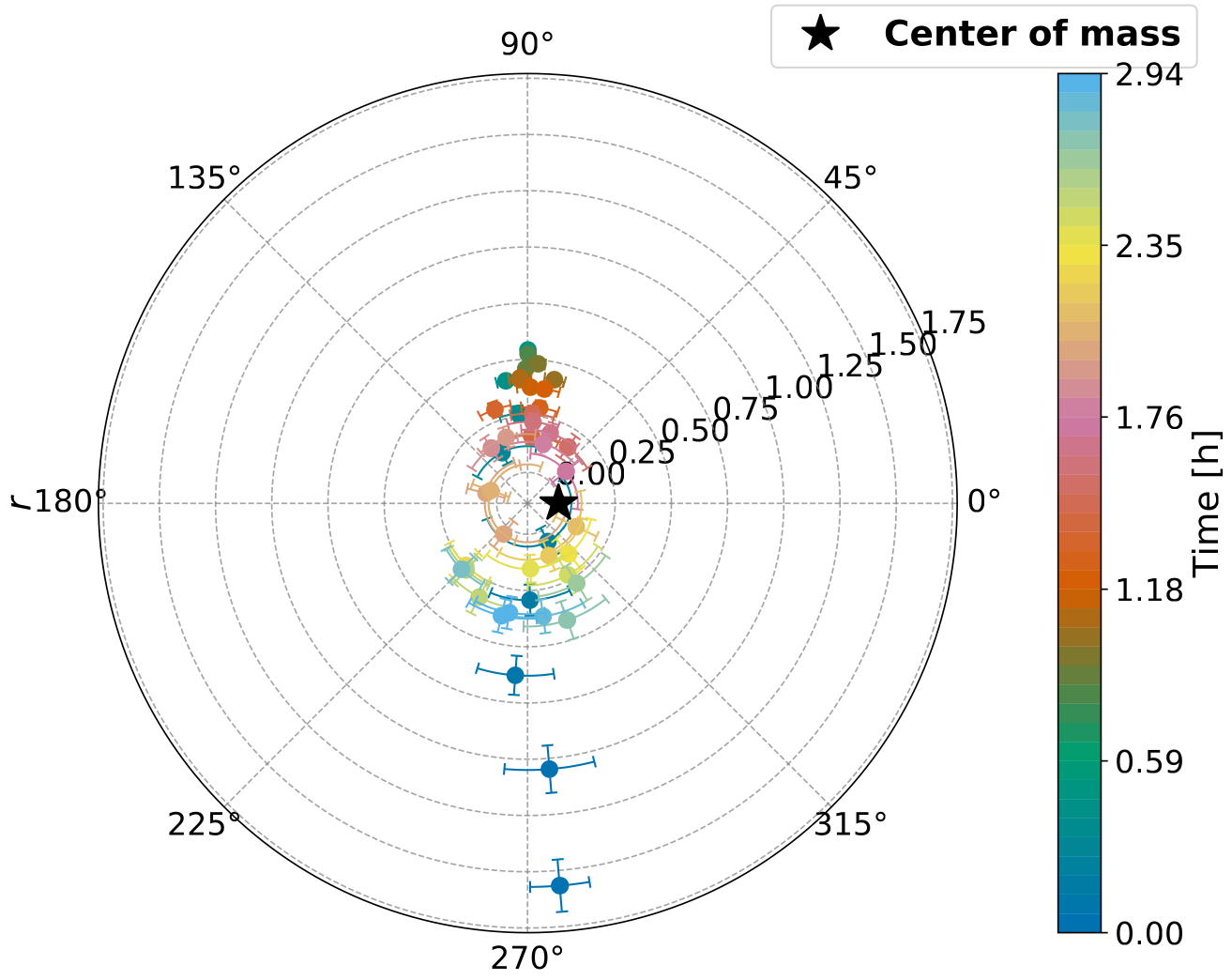}
        \label{fig:hr_flat}
    }
    \subfloat[HR diagram following $H_1$]{
        \includegraphics[width=0.32\textwidth]{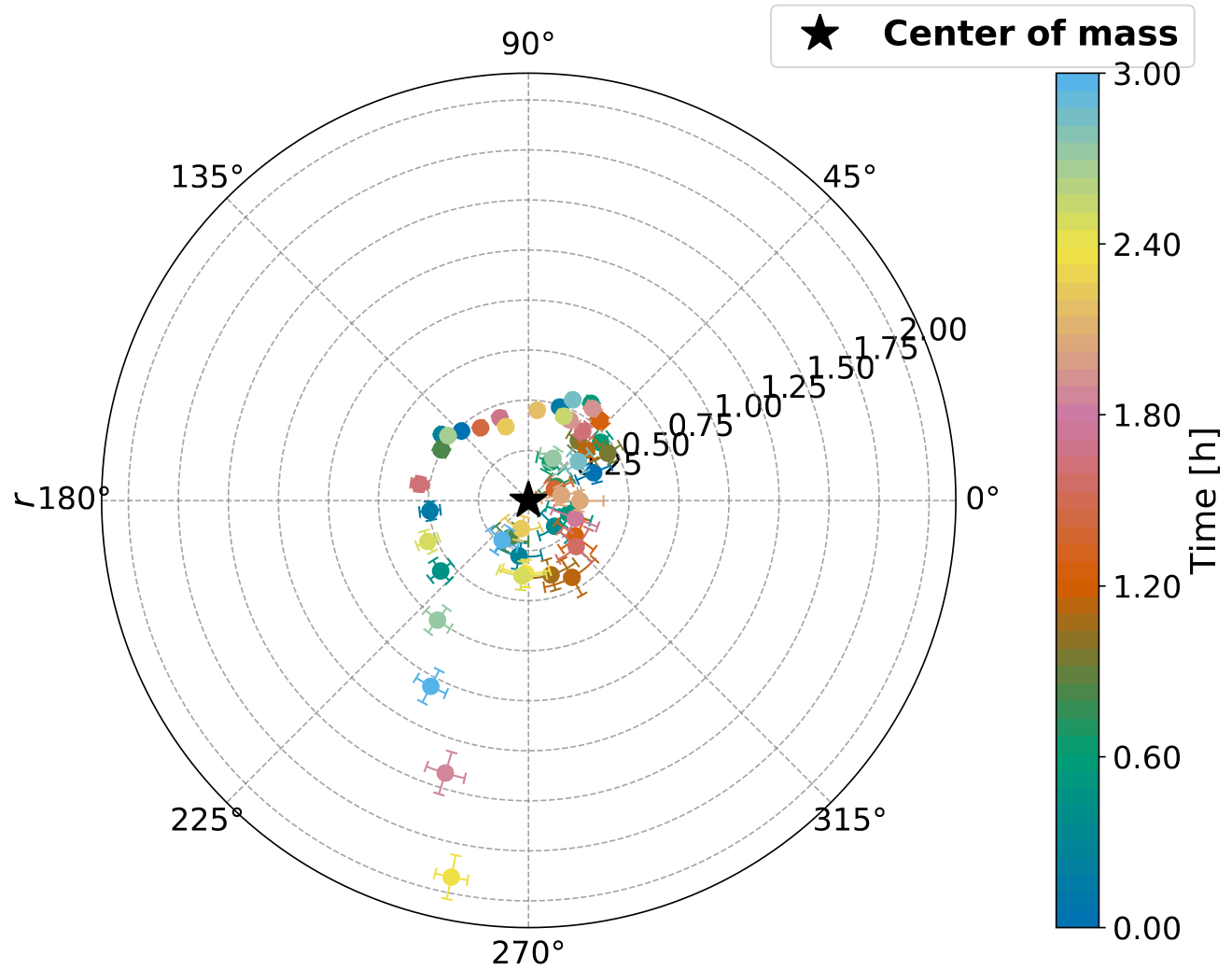}
        \label{fig:hr_loop}
    }
    \caption{Simulated HR diagrams for the three configurations described in Sec.~\ref{subsec:HR}. The color scale indicates temporal evolution. The black star is the barycenter of points in the HR diagram.}
    \label{fig:main}
\end{figure*}

In order to evaluate the distribution of $T$ for the data, we use toy MC simulations accounting for flux measurement uncertainties. Those MC simulations are used to infer the distribution of $T$ under hypotheses to be rejected. Concerning $H_0$, we generate a synthetic light-curve $\phi_{H_0}(E_2,E_3)$ that is perfectly correlated with $\phi(E_1,E_2)$. For $H_1$, we randomly shuffle the time order of $\phi(E_1,E_2)$ relative to $\phi(E_2,E_3)$. The $p$-value of the data is evaluated for both hypotheses. We convert the $p$-value into detection significance using the normal survival distribution function.
To validate this method, we simulated observed light-curves with and without injected hysteresis. The flux was simulated to increase and decrease linearly. In the hysteresis case, the rise times differ between the two bands. For the case without hysteresis, two possible scenarios are simulated, the first one with a high correlation of the flux in the two energy bands, this simulation follows the $H_0$ hypothesis. The second case is simulated assuming a "loop" signature in the HR diagram, following the $H_1$ hypothesis. Thus, these validation simulations result in three distinct data samples. The HR diagrams in polar coordinates are displayed in Fig.~\ref{fig:main}. For each sample, we applied our method to infer the $p$-values. 

\begin{figure}[t]
    \centering
    \includegraphics[width=1\linewidth]{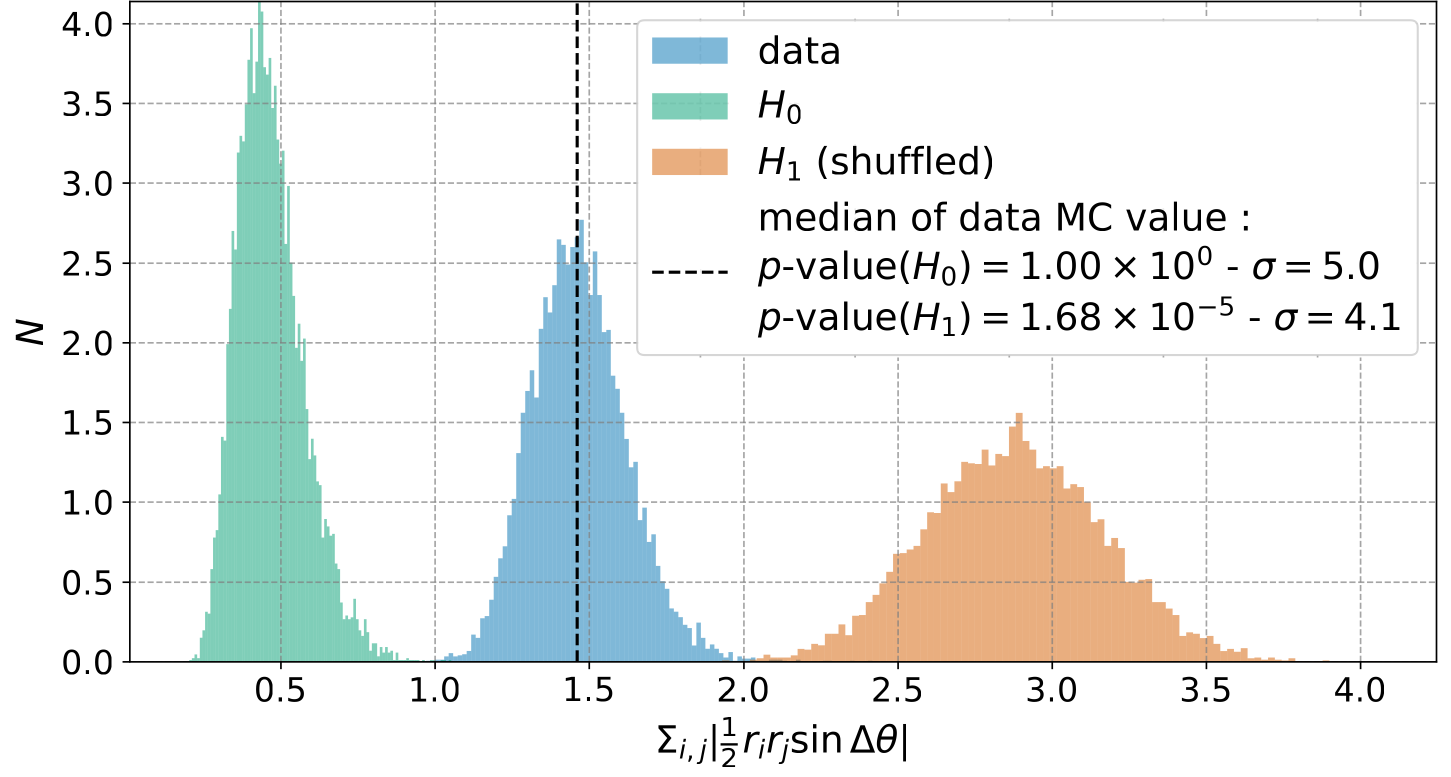}
    \caption{Distribution of $T$ for the simulated data (blue), for $H_0$ (green) and for $H_1$ (orange) hypotheses based on 10,000 MC simulations. The vertical dotted line represents the median value calculated from the simulated data with injected hysteresis. Note : the $p$-value is equal to 1 because all points of the $H_0$ distribution are lower than the observed value. The significance is then evaluated assuming a Gaussian distribution for $H_0$.}
    \label{fig:hr_dist}
\end{figure}

The distributions of $T$ for a simulation with injected hysteresis (as seen in Fig.~\ref{fig:hr}) are shown in Fig.~\ref{fig:hr_dist}. In this example, the $p$-values reject both $H_0$ and $H_1$, leading to a hysteresis detection with at least $4~\sigma$ confidence. The detection power for simulations following $H_0$ (Fig.~\ref{fig:hr_flat}) and $H_1$ (Fig.~\ref{fig:hr_loop}) is summarized in Table~\ref{tab:results_hr}. In the $H_0$ case, the estimator does not reject either hypothesis. For the $H_1$ case, the estimator highly rejects $H_0$ but remains consistent with $H_1$. 

We also tested the impact of flux uncertainties by performing simulations with uncertainties scaled by factors of 0.5 and 2. These results, presented in Table~\ref{tab:results_hr}, show that a false hysteresis detection is never confirmed when data follow $H_0$ or $H_1$. When a hysteresis is actually present, the detection confidence increases as the flux uncertainties decrease. 

In conclusion, this estimator effectively detects hysteresis by evaluating the area enclosed in the HR diagram without assuming a specific loop shape. By fully utilizing temporal information, this method provides a robust criterion for identifying spectral evolution patterns. The method has been validated here on benchmark simulations only; applying it to archival light-curves in which hysteresis has been reported, in particular in the X-ray band~\citep{2004A&A...424..841R}, is a natural further validation that we leave for future work.

\begin{table}[h]
    \centering
    \caption{Significance of hypothesis rejection (in $\sigma$) under different simulated scenarios. The nominal case corresponds to typical observation with errors scaled to about 15\% of the flux. Errors are scaled by a factor 0.5 and 2 respectively for the low and high error cases}
    \label{tab:results_hr}
    \begin{tabular}{lcccccc}
        \toprule
        & \multicolumn{2}{c}{\textbf{Low err.}} & \multicolumn{2}{c}{\textbf{Nominal err.}} & \multicolumn{2}{c}{\textbf{High err.}} \\
        \cmidrule(lr){2-3} \cmidrule(lr){4-5} \cmidrule(lr){6-7}
        Scenario simulated & \textbf{$H_0$} & \textbf{$H_1$} & \textbf{$H_0$} & \textbf{$H_1$} & \textbf{$H_0$} & \textbf{$H_1$} \\
        \midrule
        With hysteresis      & 23 & 6.8 & 5 & 4.1 & 3.3 & 0.6 \\
        Following $H_0$      & 2.1 & 0.9 & 2.1 & 1.2 & 2.1 & 1.8 \\
        Following $H_1$      & 65 & 1.6 & 22 & 0.6 & 5.2 & 0.6 \\
        \bottomrule
    \end{tabular}
\end{table}

\subsection{Search for periodicity}
\label{subsec:periodic_search}

Algorithms used to detect periodicity in \textit{Fermi}-LAT are numerous. Several of them are explained in~\citet{2020ApJ...896..134P}. In \citet{2022arXiv221101894P}, the significance of periodicity detection is defined as the median of the significance for all the methods. However for the sake of clarity, it has been decided to only focus here on one algorithm to investigate the possibility to detect periodicity with the CTAO. The Lomb-Scargle periodogram~\citep[LSP,][]{1976Ap&SS..39..447L,1982ApJ...263..835S} is the best known method for detection of periodicity in time series in astronomy. This method appears as the most appropriate because it is particularly robust for time series containing unevenly-spaced data, which is the major characteristic of observations from ground based telescopes with a relatively low duty cycle compared to space telescopes (depending of the field of view). 

To infer significance, we used only one method of~\citet{2020ApJ...896..134P}: bootstrapping. This method consists in the computation of a false alarm probability (FAP) which is associated to the detected peak in the LSP. This estimator  measures the probability that a light-curve with no signal, namely random fluctuations, generates a peak in its periodogram of similar level. FAP is a $p$-value that can be used to compute a significance associated to a peak in the LSP. Technically LSP and the bootstrap FAP are computed with \textsc{astropy} 2.0~\citep{2022ApJ...935..167A} which uses the algorithm from~\citet{2009A&A...496..577Z} implemented in \textsc{SciPy}~\citep{2020SciPy-NMeth}. To cross-check our results and to get the pre-trial confidence level of the LSP, we also implemented the bootstrap method on our end.

Another point that has to be considered is look elsewhere factors. Indeed when a LSP is computed, there is a resolution in frequency, and then, when looking for a LSP peak, a number of independent frequencies are tested. \\
The problem can also be posed differently. Rather than considering the frequency at which the peak occurs, one can fix a power threshold in the LSP and compare it with the maximum power reached in each bootstrapped periodogram. The fraction of bootstrap realisations whose maximum power exceeds the threshold gives a $p$-value that accounts for the trial factor by construction, since the null distribution is that of the largest excursion over the entire scanned frequency range. This is precisely the definition used by \textsc{astropy} when computing LSP with the bootstrapped method. 
Let $X = \{x_i,i \in [1,N]\}$ with $x_i$ the maximum power associated to the i-th LSP among $N$ bootstraps. $X$ is sorted in ascending order. Let $p$ be a power threshold in the original LSP, the $p$-value is then defined as follows : 
\begin{equation}
\label{eq:pvalue}
    p\text{-}\mathrm{value} = 1 - \frac{\mathrm{Sup}(\{i\ |\ x_i<p, x_i \in X\})}{N}
\end{equation}
The supremum is between 1 and $N$, which means that the $p$-value is between 0 and 1.

\section{Application to the blazar PG~1553$+$113 modelled with \textit{Fermi}-LAT data}
\label{section 5}

Making use of the possibilities offered with \textsc{CtaAgnVar}, a scientific study has been carried out to determine whether it will be possible to detect with the CTAO features that are currently detected only in the high energy domain with \textit{Fermi}-LAT. Beyond the demonstration of the pipeline, this study addresses a question of its own: the perspectives offered by the CTAO on the periodicity of PG~1553$+$113, and the observation strategy that such a measurement would require.

Indeed recent studies seem to hint to a periodicity detected for PG~1553$+$113 in the HE domain with \textit{Fermi}-LAT~\citep{2020ApJ...896..134P,2022arXiv221101894P}. In that respect the question whether such periodicity could also be detected with the CTAO obviously comes to mind. As mentioned in the introduction, periodicity in the gamma-ray light-curve may be related to the presence of a binary super massive black hole or geometrical effects like the jet precession~\citep{2004ApJ...615L...5R}.

It should be stressed that the existence of a periodicity at HE does not imply that a periodicity is present at VHE. The two bands are not necessarily produced in the same region of the jet, and multi-zone scenarios naturally lead to different variability patterns in the two energy ranges. In addition, no clear HE-VHE correlation has been established for this source, and no significant VHE periodicity has been found in a decade of MAGIC observations~\citep{2024MNRAS.529.3894M}. The model injected in the simulations presented below assumes, by construction, that the VHE emission follows the HE one, since it is obtained by extrapolating the \textit{Fermi}-LAT light-curve to the CTAO energy range. The results of this section must therefore be understood as the capability of the CTAO to test that hypothesis, rather than as a prediction that the periodicity is present at VHE. This is precisely the question that a simulation should address before an observation strategy is defined: knowing whether, and at which cost in observing time, the CTAO would be able to confirm or to rule out a VHE counterpart of the periodicity detected at HE is a scientific goal in itself.

\subsection{Methodology}
\label{Methodology}
To construct an AGN emission model from the \textit{Fermi}-LAT data, we extrapolate the spectrum measured by \textit{Fermi}-LAT into the CTAO energy range. First we simply extracted the light-curve from the \textit{Fermi}-LAT Light Curve Repository~\citep[LCR,][]{2023ApJS..265...31A} with a light curve with 3-day bins. The source spectrum at VHE has then been modelled with a power law with an exponential cut-off, EBL is accounted for, yielding :
\begin{equation}
    \Phi_z(E,t) = \Phi(t) \left(\frac{E}{E_0}\right)^{-\Gamma(t)} e^{-\frac{E}{E_\mathrm{cut}}} e^{-\tau_{\gamma \gamma}(E,z)}
\label{eq:spectralmodel}
\end{equation}
where $E_0$ is the reference energy, $\Phi(t)$ is the differential flux at the reference energy, $\Gamma(t)$ the photon index and $E_\mathrm{cut}$ the cut-off energy. The last factor in Equation~\ref{eq:spectralmodel} describes the absorption of VHE photons in the EBL with the optical depth $\tau(E,z)$ depending on the source's redshift, $z$, taken from the work of \citet{2011MNRAS.410.2556D} and the gamma rays energy. \\
$\Phi(t)$ and $\Gamma(t)$ are taken from the \textit{Fermi}-LAT LCR. In \citet{2020ApJ...896..134P,2022arXiv221101894P}, the periodicity is detected considering a fixed photon index, therefore in our work we also kept the photon index fixed at the value proposed in the \textit{Fermi}-LAT LCR. The synchrotron peak of PG~1553$+$113 is located at $3.89 \times 10^{15}\ \mathrm{Hz}$ in \citet{2022ApJS..263...24A}. This source is considered as a high-synchrotron-peaked blazar (HSP). The value of the energy cut-off is set as accounting for the observed correlation between the synchrotron and gamma-ray peak locations~\citep{2017MNRAS.469..255G}. Thus, we used the same hypothesis as in~\citet{2021JCAP...02..048A}. The cut-off energy $E_\mathrm{cut}$ has been set to 1 TeV.

PG~1553$+$113 is a source for which the redshift is not fully certain, mostly because of the lack of significant emission lines. Different methods are used to determine its value. Focusing on recent studies, with near-infrared and optical spectroscopy, \citet{2014A&A...570A.126L} found a lower limit of 0.30. In ultraviolet, with Lyman-$\alpha$ forest, \citet{2022MNRAS.509.4330D} found a value of 0.433. Finally, in the VHE domain, with EBL constraints \citet{2022MNRAS.511..994M} determined an upper limit of 0.48. This value of 0.48 has been selected to be the most conservative assumption.

To investigate the possibility to study and actually detect periodicity with the CTAO, the source is simulated to be observed once a week with a 30 min observation, during visible periods, from the northern site of CTAO. This observation strategy can be considered realistic because it corresponds to a reasonable time budget and provides the scheduler with greater flexibility.\\
It has been decided to perform simulation of PG~1553$+$113 observations from the 22nd of August, 2008, close to the beginning of the data taken from \textit{Fermi}-LAT, during 15 years. This strategy leads to about 20 observations per year, yielding a total of 150h for 262 observations.

\begin{figure}
    \centering
    \includegraphics[width=\hsize]{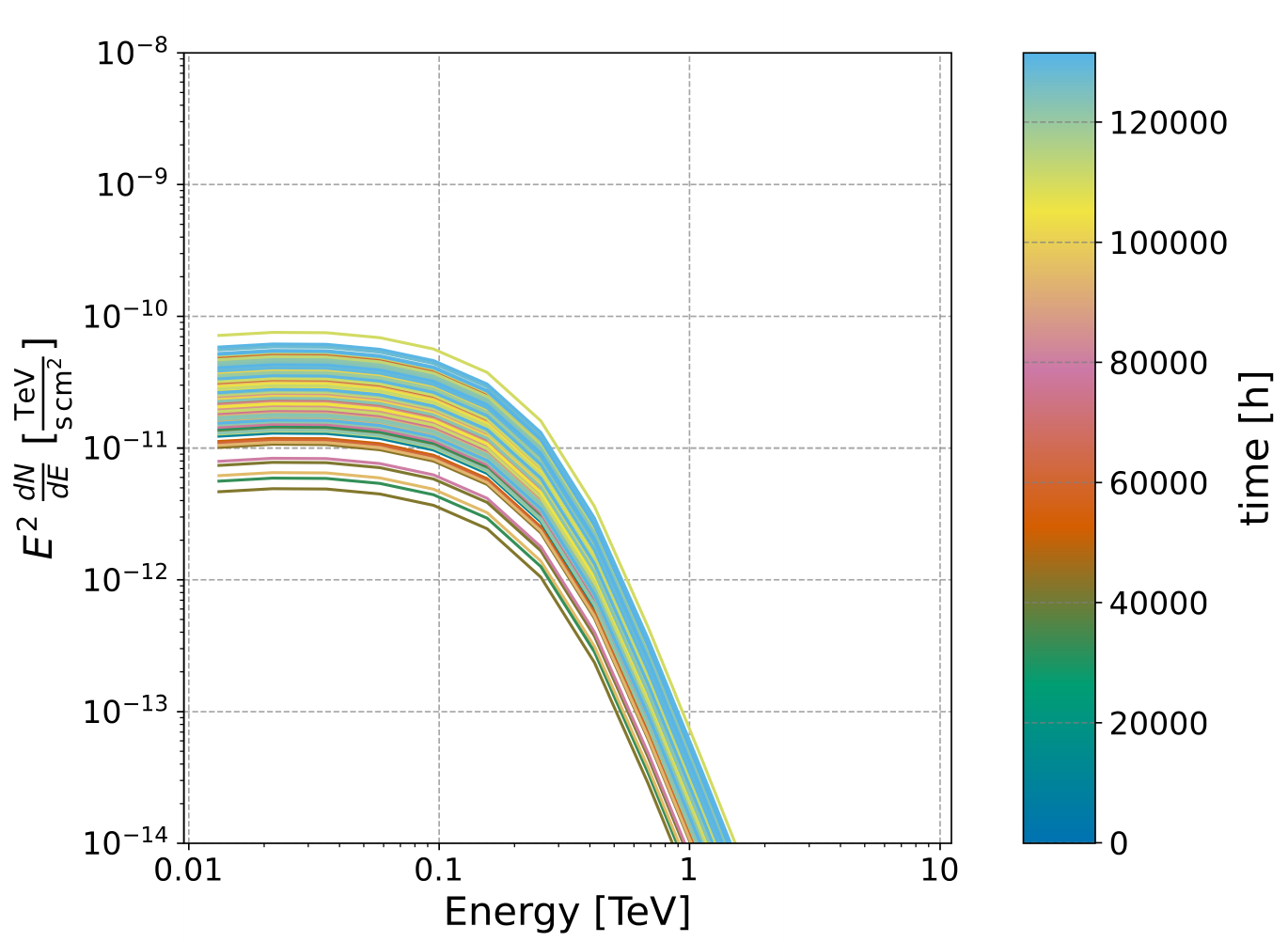}
    \caption{SED modelling used for simulations. Colors display the time evolution.}
    \label{fig:SED}
\end{figure}


\subsection{Results}

\begin{figure}
    \centering
    \includegraphics[width=\hsize]{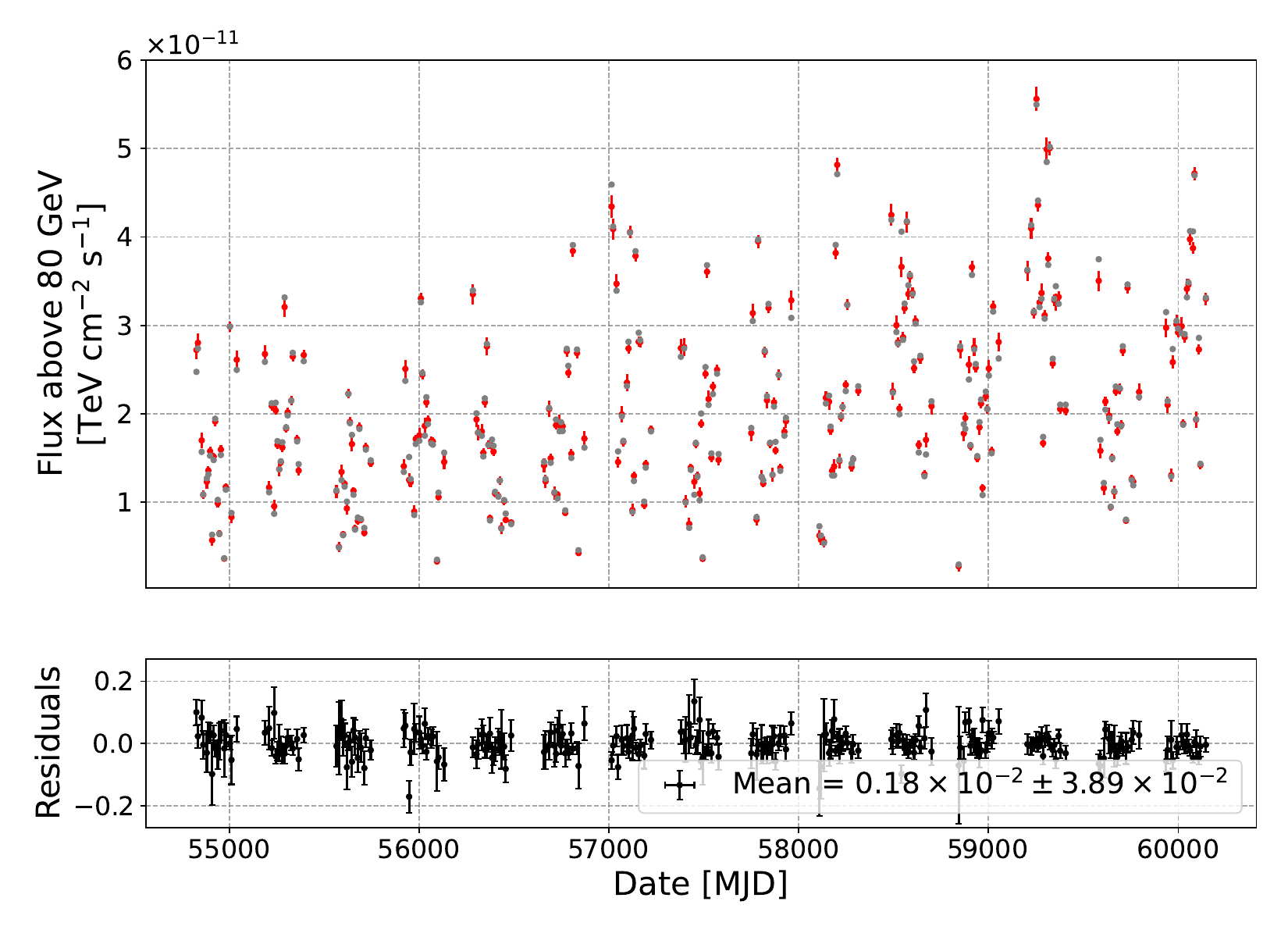}
    \caption{Light-curve reconstructed for 15 years of data observed with a weekly cadence. Red points and uncertainties are the integrated flux reconstructed between 80 GeV and 30 TeV, each point is associated to one single night. Gray points are the injected value predicted by the modelling from ~\ref{Methodology}. The residuals between data and model are shown on the lower panel.}
    \label{fig:light-curve}
\end{figure}

Simulations of 15 years of observations with a weekly cadence give the light-curve presented in Fig.~\ref{fig:light-curve}.
Over the 262 simulated 30 min observations, the mean detection significance is around 20$~\sigma$.
Looking at the residuals, the agreement between data and model is very satisfactory. As mentioned, the search for periodicity carried out in~\citep{2020ApJ...896..134P,2022arXiv221101894P} has been done with a LC evaluated with a fixed photon index. Hence, here the first analysis has been done fitting a power law EBL absorbed with free photon index, the median photon index has been evaluated to $2.15 \pm 0.05$ at 50 GeV (the injected value is 2.17). A second fit has been performed thereafter to reconstruct the light-curve presented in this work, with a photon index fixed to this value for each time bin. 

\begin{figure}
    \centering
    \includegraphics[width=\hsize]{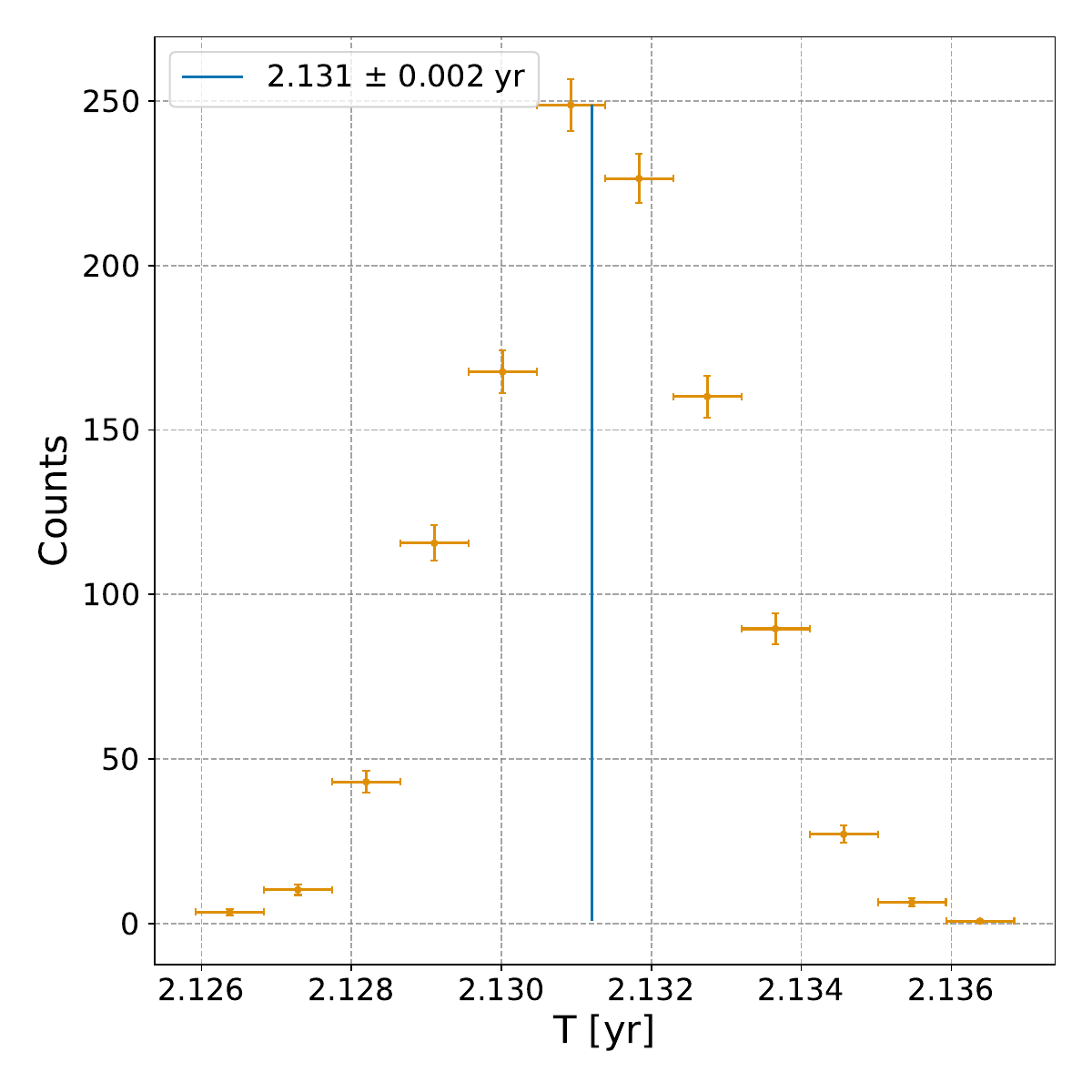}
    \caption{Histogram of the reconstructed detected periods in LSP computed with a sample of 6000 MC based on the original light-curve shown in Fig~\ref{fig:light-curve}.}
    \label{fig:period_uncertainty}
\end{figure}

As mentioned, the LSP is computed with \textsc{astropy}. The minimal period has been set to 1 year to avoid the detection of the biased period associated to moon cycles and seasonal evolution of the visibility of the source. A discussion about the impact of this choice and consequences of relaxing this limit to a lower period is done in Sec.~\ref{conclusion}. The maximal period is set to 15 years, the whole observation period. However, because of the Nyquist–Shannon limitation, we will not consider any potential detection for periods above 7.5 years. 
To estimate an uncertainty on the computed period, toy MCs have been used: a set of 6000 light-curves have been produced with points randomly chosen within the uncertainties bars of the observed light-curve. LSP have been computed for each of these light-curves and the histogram of the measured period is presented in Fig.~\ref{fig:period_uncertainty}. It should be kept in mind that this procedure propagates the statistical uncertainties of the reconstructed flux points only, the injected light-curve being a fixed function of time. The resulting dispersion therefore quantifies the reproducibility of the position of the LSP peak given the CTAO measurement uncertainties, and not the accuracy on the period of the source itself, which is set by the frequency resolution of the periodogram, of the order of $P^2/T$. The high level of accuracy in the reconstruction of the light-curve presented in Fig.~\ref{fig:light-curve} allows to keep uncertainty on the detected period very low. The detected period of PG~1553$+$113 in \citet{2020ApJ...896..134P,2022arXiv221101894P} is $2.2 \pm 0.2$ years which agrees with the value of $2.131 \pm 0.002$ years that we obtained.

\begin{figure}
    \centering
    \includegraphics[width=\hsize]{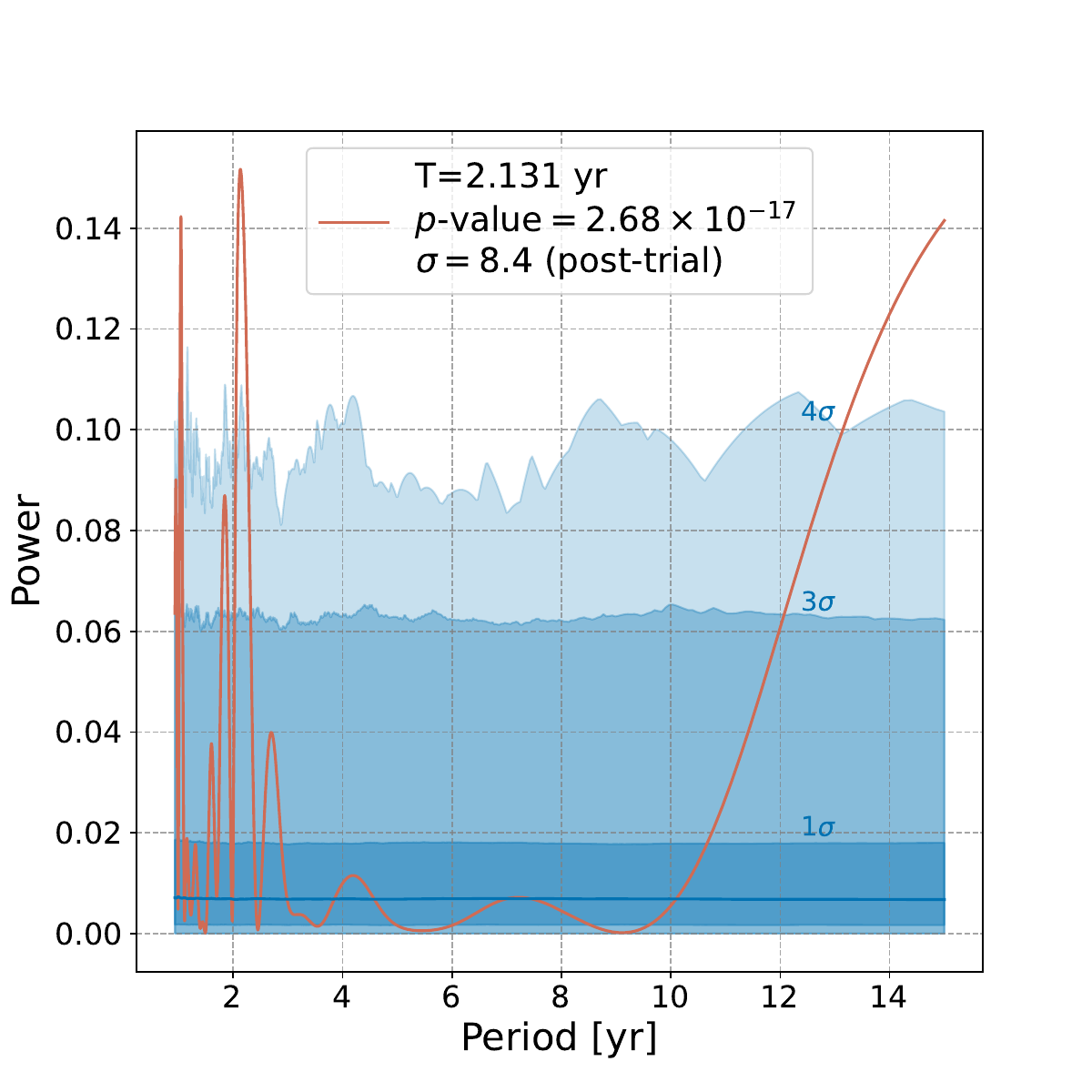}
    \caption{LSP computed with the light-curve presented in Fig.\ref{fig:light-curve}. The blue line is the median LSP computed with 500 000 bootstraps, the three blue areas are corresponding to the 1, 3 and 4 $\sigma$ (pre-trial) around the median, from the darkest to the lightest. The red line is the LSP computed on the data. The peak with the maximal power is associated to a period of $2.131$ years with a significance of $8.4~\sigma$ (post-trial).}
    \label{fig:LSP}
\end{figure}

\begin{figure*}
    \centering
    \includegraphics[width=0.9\textwidth]{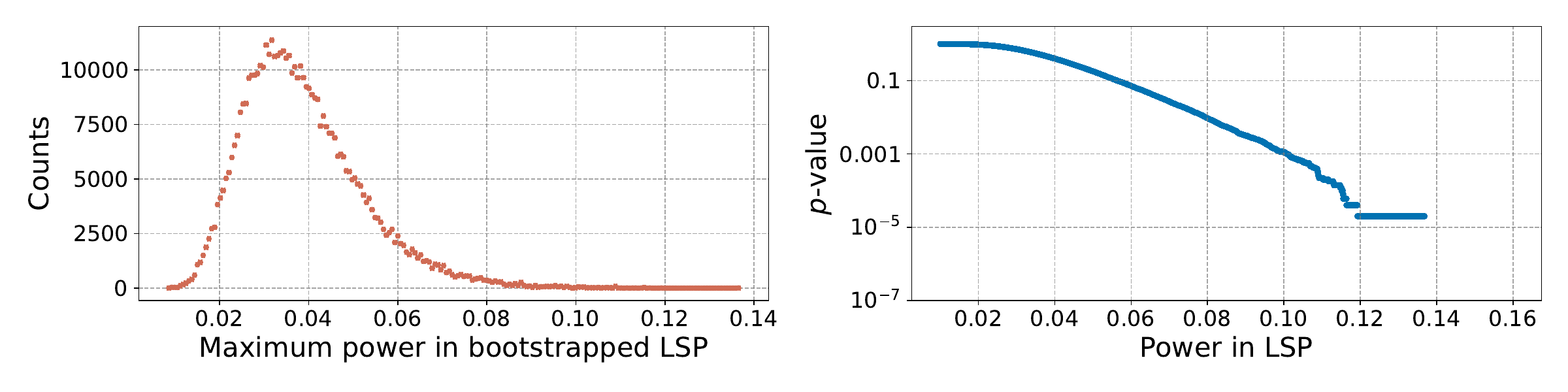}
    \caption{\textit{Left : }Distribution of the maximum power computed with the bootstrapped LSP. \textit{Right : }Evolution of the $p$-value defined in Eq.~\ref{eq:pvalue} against the power threshold in LSP. As a reminder, a value of 1 is associated to noise only. }
    \label{fig:pvalue}
\end{figure*}

Secondly, to estimate the confidence level of the detected period, 500 000 bootstrap realizations have been performed. The resulting periodogram is presented in Fig.~\ref{fig:LSP}. The peak associated to the detected period of $2.131$ years is significantly visible. To assign a significance level associated to that period, the $p$-value defined in Eq.~\ref{eq:pvalue} has been used. The distribution of the maximum power in bootstrapped LSP and the evolution of the $p$-value against the power in LSP is presented in Fig.~\ref{fig:pvalue}.
With the MC used to compute the measured period of $2.131$ years, it is also possible to estimate the value and the uncertainty of the power associated to the peak in the LSP, the value is then $0.147 \pm \ 0.002$. The $p$-value has been computed, yielding a detection significance of this peak of $8.4~\sigma$. This significance value accounts for trial factors while the blue shaded area in Fig.~\ref{fig:LSP} show the pre-trial confidence levels. 
The main caveat of the model adopted for the simulations described in this section lies in the uncertain redshift of PG~1553$+$113. Nevertheless, it should be noted that we conservatively chose an upper bound on the distance to the source. A lower redshift value would result in less EBL absorption, leading to an increased gamma-ray flux, leading to an improvement of the confidence level on the periodicity measurement. The second caveat is the one already discussed at the beginning of this section: the injected model assumes that the VHE emission follows the HE one, so that the significance quoted above is the one that would be obtained if a VHE counterpart of the HE periodicity does exist.

We can reasonably assume that with 15 years of data and weekly 30 min observations (approximately 20 observations per year), it is possible to detect the periodicity of PG~1553$+$113 with the CTAO at a significance level exceeding $5~\sigma$ (post-trial). 

To evaluate how this detection capability depends on observational parameters, we tested several alternative configurations. When reducing the observational cadence while maintaining the same total time budget, the detection significance degrades substantially. Specifically, 60 min observations every 2 weeks yield a detection at $1.5~\sigma$ (post-trial), while monthly observations of 120 min drop below $1~\sigma$ (post-trial). Conversely, increasing the cadence to 1 observation every 3 days (with 15 min duration) also decreases the significance below $1~\sigma$ (post-trial), due to poorer light-curve quality from statistical fluctuations. These results demonstrate a critical trade-off: the detection significance depends on the balance between cadence and light-curve quality.\\
Increasing the time budget by a factor of two with bi-weekly 30 min observations yields only a marginal improvement of 0.5\% in detection confidence. More significantly, extending individual observations to 60 min while maintaining weekly cadence reduces the detection to $1.5~\sigma$ (post-trial). This degradation arises because longer observations exceed the visibility windows available under good observing conditions (zenith distance constraints), reducing the total number of usable observations by approximately 30\% compared to the 30 min baseline\footnote{This reduction in significance due to visibility constraints is already present in the injected model.}.\\
Reducing the total time budget by a factor of two yields mixed results: weekly observations of 15 min fall below $1~\sigma$ (post-trial), while bi-weekly 30 min observations achieve $3~\sigma$ (post-trial).\\
Overall, these results demonstrate that weekly 30 min observations represent an optimal balance for PG~1553$+$113 between detection significance and practical observing constraints for CTAO.

The 15 years quoted above deserve to be put in perspective with the \textit{Fermi}-LAT measurement, which was obtained over a shorter baseline. The two instruments are not limited by the same factor. \textit{Fermi}-LAT operates in survey mode and covers a given source almost continuously, which allows a light-curve to be built with a regular 3-day binning. An IACT, on the contrary, is a pointing instrument whose observations are restricted by daylight, moonlight, by the range of acceptable zenith angles and by the time budget allocated to study a single source: for PG~1553$+$113 seen from the northern site, these constraints leave about 20 usable observations per year. What limits the periodicity search with the CTAO is therefore the sampling of the light-curve, and not the sensitivity of the instrument. Since the significance of a peak in the LSP grows with the number of sampled cycles, recovering a period of 2.13 years requires a baseline of the order of a decade, and the same requirement would apply to any instrument sampling the source about 20 times per year, whatever its sensitivity.

What the CTAO brings is the quality of each individual measurement. A single 30 min observation already yields a mean detection significance of about $20~\sigma$, whereas a comparable measurement requires several hours of exposure with current-generation IACTs. This precision per time bin implies that the statistical uncertainties on the flux contribute negligibly to the determination of the period, which would be limited by the sampling and by the duration of the campaign. The dispersion of $0.002$ years quoted above should therefore not be compared to the $2.2 \pm 0.2$ years reported with \textit{Fermi}-LAT, which is of the order of the periodogram resolution, here $P^2/T \approx 0.3$ years.

\section{Discussion and conclusion}
\label{conclusion}

In this article, the technical aspects of \textsc{CtaAgnVar}, a pipeline for AGN simulations and analysis with the CTAO, have been presented. Its functionalities, analysis modes and configurable options are summarized in Table~\ref{tab:functionalities}. Important features are dedicated to the study of time variability, for both short and long time scale. A statistical estimator for detecting hysteresis in an HR diagram has been derived for the first time. It has been validated on benchmark simulations and it will now be used in the CTAO extragalactic working group. The pipeline is already exploited within the CTAO Consortium for the study of blazar flares and for the definition of the long-term monitoring strategies~\citep{2023arXiv230909615C,2023arXiv230912157G}, which are the subject of forthcoming Consortium publications. \\

Simulations of PG~1553$+$113 long-term monitoring have also been performed, and constitute the second main result of this work. It has been shown that, with a modelling based on \textit{Fermi}-LAT data extrapolated to the VHE, it would be possible with the CTAO to detect the same periodicity that has been measured with \textit{Fermi}-LAT, at a $8.4~\sigma$ level (post-trial). The results presented in this paper suggest that the detection significance depends on the balance between cadence, observation duration, and the resulting number of usable observations. The optimal strategy to detect this periodicity is to perform weekly 30 min observations. The baseline of 15 years is set by the sampling imposed by the visibility of the source, about 20 observations per year, and not by the sensitivity of the instrument: each 30 min observation already reaches a mean significance of about $20~\sigma$, so that the determination of the period would be limited by the sampling of the light-curve and not by the photon statistics, while a decade of MAGIC observations did not lead to a significant detection of this periodicity~\citep{2024MNRAS.529.3894M}.\\

One important limitation to discuss concerns the minimum detectable period, set to one year in this work. CTAO observations are inherently constrained by lunar cycles and the seasonal evolution of source visibility. These observational constraints introduce systematic periodicities on monthly-to-yearly timescales. Consequently, noise in the LSP for periods below one year can dominate, reducing the significance of the periodicity detection as computed through bootstrapping, with an estimated reduction of 15\% to 30\%. Therefore, searching for periodicities below one year with the CTAO remains challenging for sources fainter than PG~1553$+$113 using the method presented in this work. Conversely, for periods longer than one year, the method is expected to provide sufficient sensitivity.\\

As mentioned this pipeline is currently widely used within the CTAO Consortium to study CTAO prospects in terms of AGN variability. Its use is however not restricted to the CTAO nor to the CTAO Consortium. The instrument enters the pipeline only through the IRFs and the geographical location of the array, so that the IRFs of any other current or future IACT can be used instead. Being open-source, \textsc{CtaAgnVar} can therefore be used by the community at large to investigate the capabilities of current and future IACTs regarding AGN variability. \textsc{CtaAgnVar} has been built to study time-dependent phenomena. It can therefore be efficient to simulate data of other transient phenomena, such as gamma-ray bursts. Investigation of the violation of Lorentz invariance with AGN also started recently with \textsc{CtaAgnVar}.
Another important point to notice is that this pipeline is both able to simulate data and to analyse real data. In this scope, analyses of data from the first LST-1 of the CTAO have started.

\begin{acknowledgements}

This work has been carried out in close collaboration with the Extragalactic Working Group of the CTAO Consortium. We specifically acknowledge members of the AGN flare and AGN long-term monitoring task forces, whose feedback has been invaluable in establishing the results presented here.

This work made use of Gammapy \citep{gammapy:2023}, a community-developed Python package. The Gammapy team acknowledges all Gammapy past and current contributors, as well as all contributors of the main Gammapy dependency libraries: \href{https://numpy.org/}{NumPy}, \href{https://scipy.org/}{SciPy}, \href{http://www.astropy.org}{Astropy}, \href{https://astropy-regions.readthedocs.io/}{Astropy Regions}, \href{https://scikit-hep.org/iminuit/}{iminuit}, \href{https://matplotlib.org/}{Matplotlib}.

\end{acknowledgements}

\bibliographystyle{aa} 
\bibliography{CtaAgnVar}

\end{document}